# Two-stage Planning for Electricity-Gas Coupled Integrated Energy System with CCUS Considering Carbon Tax and Price Uncertainties

Ang Xuan, *Student Member, IEEE,* Xinwei Shen, *Senior Member, IEEE,*
Qinglai Guo, *Senior Member, IEEE,* Hongbin Sun, *Fellow, IEEE*

*Abstract*—In this article, we propose two-stage planning models for Electricity-Gas Coupled Integrated Energy System (EGC-IES), in which traditional thermal power plants (TTPPs) are considered to be retrofitted into carbon capture power plants (CCPPs), with power to gas (PtG) coupling CCPPs to gas system. The sizing and siting of carbon capture, utilization and storage (CCUS)/PtG facilities, as well as the operation cost of TTPPs/CCPPs/gas sources/PtG, are all considered in the proposed model, including penalty on carbon emissions and revenue of CCUS. With changing policy on climate change and carbon emission regulation, the uncertainties of carbon price and carbon tax are also analyzed and considered in the proposed planning model. The stochastic planning, and robust planning methods are introduced to verify mutually through economic and carbon indices. The proposed methods' effectiveness in reducing carbon emissions, increasing CCUSs' profit from EGC-IES are demonstrated through various cases and discussions.

## Nomenclature

**Abbreviations:**

| | |
|---|---|
| CCUS | carbon capture, utilization and storage |
| CC | carbon capture |
| CS | carbon storage |
| IEA | International Energy Agency |
| TTPP | traditional thermal power plant |
| CCPP | carbon capture power plant |
| EGC-IES | Electricity-Gas Coupled Integrated Energy System |
| PtG | Power to gas |

**Sets:**

| | |
|---|---|
| $\Psi^{GN}$ | Set of gas nodes |
| $\Psi^{GP}$ | Set of gas pipelines |
| $\Psi^{GS}$ | Set of natural gas sources |
| $\Psi^{EB}$ | Set of electric buses |
| $\Psi^{TL}$ | Set of transmission lines |
| $\Psi^{GEN}$ | Set of electric generators |
| $\Psi^{CCPP}$ | Set of carbon capture power plants |

**Indices:**

| | |
|---|---|
| $t$ | Index of time period |
| $p$ | Index of gas pipeline |
| $l$ | Index of electricity transmission line |
| $m$ | Index of gas node |
| $n$ | Index of electricity node |
| $i$ | Index of natural gas source |
| $j$ | Index of electric generator |
| $k$ | Index of piecewise linearization segment |

**Parameters:**

| | |
|---|---|
| $T$ | Time horizon, hour |
| $\kappa$ | Annualized coefficient of investment cost, dimensionless |
| $dr$ | Discount rate, dimensionless |
| $L^\tau$ | Lifetime of facility $\tau$, year |
| $W_p$ | Weymouth coefficient of pipeline $p$, Mm$^3$/(h·bar) |
| $\overline{\pi}_m / \underline{\pi}_m$ | Maximum/minimum nodal pressure of gas node $m$, bar |
| $seg$ | Number of piecewise linearization segment, dimensionless |
| $F^{gas}_{p,k}$ | Lowest boundary point of gas flow in segment $k$ on pipeline $p$, Mm$^3$/h |
| $\overline{P}_i / \underline{P}_i$ | Maximum/minimum outputs power of gas source $i$, Mm$^3$/h |
| $\Delta \overline{P}_i$ | Maximum ramp up power of gas source $i$, Mm$^3$/h |
| $L_{m,t}$ | Gas load of gas node $m$ in hour $t$, Mm$^3$/h |

Ang Xuan and Xinwei Shen are with Tsinghua-Berkeley Shenzhen Institute, Tsinghua Shenzhen International Graduate School, Tsinghua University. Qinglai Guo and Hongbin Sun are with Dept. of Electrical Engineering, Tsinghua University. This work is supported by the National Natural Science Foundation of China (No. 52007123)
(Corresponding author: Xinwei Shen, email: sxw.tbsi@sz.tsinghua.edu.cn; Hongbin Sun, email: shb@tsinghua.edu.cn).

| | |
|---|---|
| $X_l$ | Reactance of transmission line $l$, Ω |
| $\overline{F}_l^{ele}$ | Maximum power flow of transmission line $l$, MW |
| $T_j^{on}/T_j^{off}$ | Maximum warm-up/cool-down time of generator $j$, hour |
| $emi$ | CO$_2$ emission factor of TTPP, ton/MW |
| $W^{CC}$ | CCPP energy consumption of capturing per ton CO$_2$, MWh/ton |
| $\alpha^{CO_2}$ | CO$_2$ consumed volume of PtG to generate per unit CH$_4$, ton/MWh |
| $\eta^{PtG}$ | Conversion efficiency of PtG, dimensionless |

*Matrixes:*

| | |
|---|---|
| **A** | Gas node-gas source incidence matrix |
| **B** | Gas node-gas pipeline incidence matrix |
| **C** | Bus-generator incidence matrix |
| **D** | Bus-branch incidence matrix |

*Variables:*

| | |
|---|---|
| $y_j^{PtG}$ | Integer variables, PtG planned capacity retrofitting from generator $j$ |
| $s_{m,j}$ | Binary variables, $s_{m,j}=1$ if gas node $m$ is connected to PtG $j$ |
| $P$ | Operation power, MW |
| $Q$ | CO$_2$ volume, ton |
| $f_{p,t}^{gas}$ | Gas flow on pipeline $p$ in hour $t$, Mm$^3$/h |
| $\pi_{m,t}$ | Nodal pressure of gas node $m$ in hour $t$, bar |
| $\phi_{p,t,k}$ | Continuous variable of segment $k$ on pipeline $p$ in hour $t$ indicating the proportion occupied in a specific segment |
| $\delta_{p,t,k}$ | Binary variable of segment $k$ on pipeline $p$ in hour $t$ indicating the selected status of that segment |
| $f_{l,t}^{ele}$ | Power flow on transmission line $l$ in hour $t$, MW |
| $\theta_{n,t}$ | Phase angle of electric node $n$ in hour $t$ |
| $u_{j,t}$ | Binary variable indicating start-stop status |
| $v_{j,t}/w_{j,t}$ | Binary variable reflecting startup/shutdown action |

## I. INTRODUCTION

### A. Background and Motivation

THE emission of greenhouse gas which gives priority to carbon dioxide (CO$_2$), is a primary driver of global climate change and one of the most pressing challenges nowadays [1]. In response to global climate change, the Paris Agreement was adopted by 196 parties at the 21st Conference of the United Nations Framework Convention on Climate Change in Paris on 12 Dec., 2015 and entered into force on 4 Nov., 2016. Its long-term goal is to keep the rise in global average temperature to well below 2°C above pre-industrial levels, and to pursue efforts to limit the increase to 1.5°C [2].

It has already sparked various low-carbon solutions since the Paris Agreement entered into force, more and more countries, regions and cities all over the world are establishing zero-carbon targets. Carbon capture, utilization and storage (CCUS) is the key technology that reduces and removes CO$_2$ after emission, which is a critical part of zero-carbon goals [3]. According to Special Report on CCUS [4] released by International Energy Agency (IEA), the next decade will be critical to the zero-carbon emission goal via retrofitting existing power and industrial facilities. Thus carbon capture power plant (CCPP), which is altered from traditional thermal power plant (TTPP), its carbon reduction performance in Gas Network, Electricity Network and even Electricity-Gas Coupled Integrated Energy System (EGC-IES) are worth investigating urgently.

Encouragingly, as more ambitious climate pledges are taken, many programs and strategies are factoring in the role and potential for carbon pricing and carbon markets [5]. How to translate commitment into reality in ensuring we can confine global warming to below 2°C, the careful planning for EGC-IES considering the global changing and divergent carbon price policy will be critical.

### B. Literature Review

Several scholars have carried out related researches in developing planning models considering CCUS. In CCUS and its applications, research [6] introduced a scalable infrastructure model to determine where and how much CO$_2$ to capture and store, where to build and connect pipelines. A mixed integer linear programming (MILP) model was developed [7] to describe a general modeling approach for optimal planning of energy systems subject to carbon and land footprint constraints. Several literatures addressed the retro fitment [8], expansion ([9], [11]), transition [10] pipeline networks ([12], [13]) planning problems considering CCUS.

Later in 2016, literature [14] presented a planning framework considering CO$_2$ supply, production, transportation, and emission costs. literature [15] developed a two-step linear optimization to make a trade-off between the cost of network modification and CO$_2$ emissions considering the network revamp. In recent years, CO$_2$ utilization has attracted more and more attention from scholars to industry. Review [16] and [17] summarized the existing and under-development technologies for CO$_2$ utilization in the world. The exact applications in gas field [18], oil field [19], and chemical products [20] were gradually studied in depth. CCUS is often planned with particular actual cases to observe performance. For example, a multi-stage mixed integer programming (MIP) model for CCUS planning was developed and tested in Beijing–Tianjin–Hebei, China [21], the United States [22] and the United Kingdom [23].

Specializing on CCUS's application in retrofitting existing facilities, early research [24] presented linear models of the most common components in the CCPP. A post-combustion and solvent/sorbent separation technology based CCPP model is established in [25], the authors investigated the performance of CCPP in the carbon emission market [26] in 2012. The TTPP was equipped with nuclear units, renewable energy units, and different fossil fuel-fired units in [27] to form MINLP and solve using particle swarm optimization algorithm.

Table I categories these researches by topics and mathematical method for CCUS planning in detail.

Table I Summary of Planning Model Considering CCUS

|  | Category | Literature |
|---|---|---|
| Topics | Carbon capture & storage | [6]-[11] |
|  | Carbon transportation (pipeline planning) | [12]-[15] |
|  | Carbon utilization | [16]-[23] |
| Method | Decision tree method | [13] |
|  | MILP | [6]-[8],[12],[14],[15],[24],[25] |
|  | MINLP | [9]-[11],[26]-[27] |

### C. Problem Identification and Main Contributions

Due to the long time horizons involved in CCUS planning, it is necessary to include uncertainties. Several literatures have studied relevant researches on uncertainties such as load and renewable generation uncertainties [28], generation and demand side uncertainties [29], user behavior uncertainty [30], etc. However, as the significant contents of low-carbon development, the volatilities of carbon price and carbon tax are essential sources of uncertainties that cannot be ignored.

Besides, gas-fired power generation units are considered as the coupling point of gas and electricity systems [31] for a long time. The development of power to gas (PtG) technology could be a new coupling point [33] in the future. However, related research in EGC-IES is still very few.

In this paper, two-stage planning methods for EGC-IES with CCUS considering carbon price uncertainty are proposed. Main contributions of the paper are therefore three-fold:

(1) A CCPP planning model retrofitted from TTPP, with PtG in EGC-IES for $CO_2$ utilization is proposed, in which the methane produced by PtG is transported to the gas network to relieve the gas supply pressure.

(2) The carbon penalty and carbon revenue are introduced in the objective function considering carbon tax and carbon price uncertainties globally.

(3) Three planning models aim at finding out a balance point where the external revenue and reduced penalty to cover the investment cost. The effectiveness of the proposed model is also verified.

The remainder of this paper is organized as follows: Section II would outline the CCUS business model under carbon market and carbon policy. Model formulation is displayed in Section III where the detailed objective function and constraints are described. Section IV discusses the proposed four models. Case studies and conclusions are given in Section V and VI.

## II. PRELIMINARY OF CCUS AND ITS BUSINESS MODEL

CCUS is a crucial emissions reduction technology that can be applied across the energy system [3]. Fig. 1 illustrates the significant profits method of CCUS currently. Its business model can be summarized as follows[4]: CCUS involve $CO_2$ capture from fuel combustion or industrial processes, uncaptured $CO_2$ is subject to the mandatory penalty in the form of carbon tax; captured $CO_2$ is transported via vehicles or pipeline or permanently stored deep underground, $CO_2$ could be utilized as a resource to create valuable products and services such as $CO_2$ enhanced oil recovery, $CO_2$ enhanced coal bed methane, chemical and biological products; in an emission trading system, captured $CO_2$ can also be traded, as predicted in *The New York Times* that carbon will be the world's biggest commodity market [32].

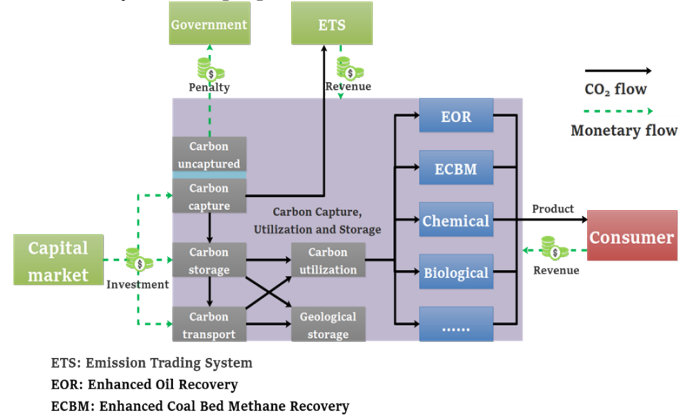

ETS: Emission Trading System
EOR: Enhanced Oil Recovery
ECBM: Enhanced Coal Bed Methane Recovery

Fig. 1 CCUS business model flow chart

Nowadays, there are 22 CCUS facilities worldwide with the capacity to capture more than 40 Mt $CO_2$ every year. Report [4] pointed out three aspects that summarize the growth trends in CCUS projects over the following decades:

*1) Retrofitting of existing power and industrial facilities that significantly reduce emissions.*
*2) The scale-up of low-carbon hydrogen production with CCUS.*
*3) The rapid adoption of CCUS technologies and applications that are not yet widely used.*

According to the technical development level, the current focus of CCUS is on retrofitting fossil fuel-based power and industrial plants.

Due to the vast differences in the economic and technological development levels and resource endowments around the world, this article uses the technical parameters published by IEA and related literatures to build a general planning model for TTPP retrofitting into CCPP, other than focus on one or several specific CCUS technologies; at the same time, based on the consideration of carbon policies are influenced by politics, culture and other factors, carbon price and carbon tax are regarded as uncertainties, the business models like carbon penalty and carbon revenue are adopted to form the planning-operation two-stage model.

## III. MODEL FORMULATION

In Section III, the objective function, gas network model, electricity network model, CCPP and PtG coupling model, and constraints on facility investment and siting are elaborated separately.

### A. Objective function

The objective function is to minimize the total cost (1), which comprises 1) investment cost $C_{inv}$ of PtGs and their sittings(gas pipelines) in gas network (2a)-(2b); 2) operation cost of gas sources $C_{ope}^{GS}$ (3a), generators $C_{ope}^{GEN}$ (3b), and PtGs $C_{ope}^{PtG}$ (3c); 3) carbon capture cost $C^{CC}$ (4a) and carbon storage cost $C^{CS}$ (4b); 4) total carbon penalty $C_{pen}$ (5a) and total carbon revenue $C_{tra}$ (5b).

$y_j^{PtG}$ and $s_{m,j}$ are planning decision variables denoting PtG sizing and siting, which are discussed in Section III.E in detail. $c_{inv}^{PtG}, c_{inv}^{siting}$ represent unit investment cost of PtG and siting, the investment costs are annualized with amortization coefficient $\kappa^{PtG}, \kappa^{siting}$ over their lifetime. $r_i, r_j, r_q$ denote corresponding operation cost of gas source $i$, generator $j$ and PtG $q$, $r^{CC}, r^{CS}$ represent carbon capture cost and carbon storage cost per ton of $CO_2$, $r^{tax}, r^{pr}$ are the carbon tax and carbon price (unit: \$/ton). For operation decision variables, $P_{i,t}, P_{j,t}, P_{q,t}$ denote operation power of gas source $i$, generator $j$ and PtG $q$ in hour $t$, $Q_{j,t}^{EMI}, Q_{j,t}^{CC}, Q_{j,t}^{CS}$ are $CO_2$ volume indicating emission, capture and storage of CCPP $j$ in hour $t$, respectively.

$$\min \ obj = \begin{pmatrix} C_{inv} + C_{ope}^{GS} + C_{ope}^{GEN} + C_{ope}^{PtG} + \\ C^{CC} + C^{CS} + C_{pen} + C_{tra} \end{pmatrix} \quad (1)$$

$$C_{inv} = \kappa^{PtG} c_{inv}^{PtG} \sum_j y_j^{PtG} + \kappa^{siting} c_{inv}^{siting} \sum_m \sum_j s_{m,j},$$
$$\forall j \in \Psi^{CCPP}, \forall m \in \Psi^{GN} \quad (2a)$$

$$\kappa^\tau = \frac{dr(1+dr)^{L^\tau}}{(1+dr)^{L^\tau} - 1}, \tau = \{PtG, siting\} \quad (2b)$$

$$C_{ope}^{GS} = \sum_i r_i \sum_t P_{i,t}, \forall i \in \Psi^{GS}, \forall t \in [1,T] \quad (3a)$$

$$C_{ope}^{GEN} = \sum_j r_j \sum_t P_{j,t}, \forall j \in \Psi^{GEN}, \forall t \in [1,T] \quad (3b)$$

$$C_{ope}^{PtG} = \sum_q r_q \sum_t P_{q,t}, \forall q \in \Psi^{PtG}, \forall t \in [1,T] \quad (3c)$$

$$C^{CC} = r^{CC} \sum_j \sum_t Q_{j,t}^{CC}, \forall j \in \Psi^{CCPP}, \forall t \in [1,T] \quad (4a)$$

$$C^{CS} = r^{CS} \sum_j \sum_t Q_{j,t}^{CS}, \forall j \in \Psi^{CCPP}, \forall t \in [1,T] \quad (4b)$$

$$C_{pen} = r^{tax} \sum_j \sum_t (Q_{j,t}^{EMI} - Q_{j,t}^{CC}), \forall j \in \Psi^{CCPP}, \forall t \in [1,T] \quad (5a)$$

$$C_{tra} = r^{pr} \sum_j \sum_t Q_{j,t}^{CS}, \quad \forall j \in \Psi^{CCPP}, \forall t \in [1,T] \quad (5b)$$

### B. Gas network model

A typical gas network model comprises natural gas transmission pipelines, natural gas sources and gas loads. The Weymouth equation [34] is adopted in (6) to describe natural gas transmission flow in this model, in which the gas flow is expressed as a quadratic equation of nodal gas pressure, $a$, $b$ denote the input end and output node respectively, in accordance with the specified direction. Formula (7a) introduces the variable $I$ to replace nodal pressure $\pi$ to avoid the nonconvexity.

$$f_{p,t}^{gas} |f_{p,t}^{gas}| = W_p^2 (\pi_{a,t}^2 - \pi_{b,t}^2), \forall p \in \Psi^{GP}, \forall t \in [1,T] \quad (6)$$

$$f_{p,t}^{gas} |f_{p,t}^{gas}| = W_p^2 (I_{a,t} - I_{b,t}), \forall p \in \Psi^{GP}, \forall t \in [1,T] \quad (7a)$$

$$\underline{\pi}_m^2 \leq I_{m,t} \leq \overline{\pi}_m^2, \forall m \in \Psi^{GN}, \forall t \in [1,T] \quad (7b)$$

Piecewise linearization are listed as constraints (8a)-(10), adopted from [35] with continuous variable $\delta$ and binary variable $\phi$: the former indicating the proportion occupied in a specific segment and the latter indicating the selected status of the segment (the value of $\phi$ changing from 1 to 0 means the segment is selected). $F_{p,k}^{gas}$ represents the lowest boundary point of gas flow in segment $k$ on pipeline $p$.

$$\phi_{p,t,k} \leq \delta_{p,t,k}, \forall p \in \Psi^{GP}, \forall t \in [1,T], \forall k \in [1, seg-1] \quad (8a)$$

$$0 < \delta_{p,t,k} \leq 1, \forall p \in \Psi^{GP}, \forall t \in [1,T], \forall k \in [1, seg] \quad (8b)$$

$$\delta_{p,t,k+1} \leq \phi_{p,t,k}, \forall p \in \Psi^{GP}, \forall t \in [1,T], \forall k \in [1, seg-1] \quad (8c)$$

$$f_{p,t}^{gas} = F_{p,1}^{gas} + \sum_{k=1}^{seg} \delta_{p,t,k} (F_{p,k+1}^{gas} - F_{p,k}^{gas}),$$
$$\forall p \in \Psi^{GP}, \forall t \in [1,T], \forall k \in [1, seg-1] \quad (9)$$

$$W_p^2 (I_{a,t} - I_{b,t}) =$$
$$F_{p,1}^{gas} |F_{p,1}^{gas}| + \sum_{k=1}^{seg} \delta_{p,t,k} (F_{p,k+1}^{gas} |F_{p,k+1}^{gas}| - F_{p,k}^{gas} |F_{p,k}^{gas}|) \quad (10)$$
$$\forall p \in \Psi^{GP}, \forall t \in [1,T], \forall k \in [1, seg-1]$$

Natural gas source production is limited by output constraints (11a) and ramp constraints (11b). Eq. (12) is the power balance constraint in gas network model among gas source, gas flow, and gas load, where on right-hand side $L_{m,t}$ is the gas load at node $m$ to be balanced, $\mathbf{A}_{m,i}$ is the element of gas node-gas source incidence matrix $\mathbf{A}$, $\mathbf{A}_{m,i}=1$ if gas source $i$ is connected to gas node $m$, $\mathbf{B}_{m,p}$ is the element of gas node-gas pipeline incidence matrix $\mathbf{B}$, $\mathbf{B}_{m,p}=1$ equals 1 if pipeline $p$ starts at node $m$, $\mathbf{B}_{m,p}=-1$ if pipeline $p$ ends at node $m$, otherwise their values are 0.

$$\underline{P}_i \leq P_{i,t} \leq \overline{P}_i, \forall i \in \Psi^{GS}, \forall t \in [1,T] \quad (11a)$$

$$-\Delta \overline{P}_i \leq P_{i,t} - P_{i,t-1} \leq \Delta \overline{P}_i, \forall i \in \Psi^{GS}, \forall t \in [2,T] \quad (11b)$$

$$\mathbf{A}_{m,i} P_{i,t} + \mathbf{B}_{m,p} f_{p,t}^{gas} = L_{m,t}$$
$$\forall m \in \Psi^{GN}, \forall i \in \Psi^{GS}, \forall p \in \Psi^{GP}, \forall t \in [1,T] \quad (12)$$

## C. Electricity network model

A typical electricity power system comprises electricity transmission lines, generators, and electricity loads. The DC power flow model in eq. (13a)-(13b) can be adopted to estimate steady state for the distribution system, (14) is constraint for power flow on transmission line.

$$f_{l,t}^{ele} = (\theta_{a,t} - \theta_{b,t})/X_l, \forall l \in \Psi^{TL}, \forall t \in [1,T] \quad (13a)$$

$$-\pi \leq \theta_{n,t} \leq \pi, \forall n \in \Psi^{EB}, \forall t \in [1,T] \quad (13b)$$

$$-\overline{F}_l^{ele} \leq f_{l,t}^{ele} \leq \overline{F}_l^{ele}, \forall l \in \Psi^{TL}, \forall t \in [1,T] \quad (14)$$

Generators' output constraints (15) and ramp constraints (16) of are similar to those of natural gas sources, in which $u$ indicating the start-stop status, the value of $u$ keeps 1 if the generator is in operation, otherwise being 0. The unit commitment constraints in formulas (17a)-(19) are employed to accurately describe generator output characteristics. $v$ and $w$ are binary variables to reflect startup and shutdown actions, the value of $v$ is 1 if the generator started up from the prior moment, and the value of $w$ is 1 if the generator shut down from the prior moment, otherwise they remain 0.

$$u_{j,t}\underline{P}_j \leq P_{j,t} \leq u_{j,t}\overline{P}_j, \forall j \in \Psi^{GEN}, \forall t \in [1,T] \quad (15)$$

$$-\Delta \overline{P}_j \leq P_{j,t} - P_{j,t-1} \leq \Delta \overline{P}_j, \forall j \in \Psi^{GEN}, \forall t \in [2,T] \quad (16)$$

$$\sum_{tt=t-T_j^{on}+1}^{t} v_{j,tt} \leq u_{j,t}, \forall j \in \Psi^{GEN}, \forall t \in [T_j^{on}, T] \quad (17a)$$

$$\sum_{tt=t-T_j^{off}+1}^{t} w_{j,tt} \leq 1-u_{j,t}, \forall j \in \Psi^{GEN}, \forall t \in [T_j^{off}, T] \quad (17b)$$

$$u_{j,t} - u_{j,t-1} = v_{j,t} - w_{j,t}, \forall j \in \Psi^{GEN}, \forall t \in [2,T] \quad (18)$$

$$v_{j,t} + w_{j,t} \leq 1, \forall j \in \Psi^{GEN}, \forall t \in [1,T] \quad (19)$$

Eq. (20) is the power balance constraint in electricity network among generator, power flow, and load, similar to eq. (12).

$$\mathbf{C}_{n,j}P_j + \mathbf{D}_{n,l}f_{l,t}^{ele} = L_{n,t}$$
$$\forall n \in \Psi^{EB}, \forall j \in \Psi^{GEN}, \forall l \in \Psi^{TL}, \forall t \in [1,T] \quad (20)$$

## D. CCPP and PtG coupling model

A coupling model combined TTPP, CC, CS and PtG technology to retrofit into CCPP is illustrated below, where exhaust emission from TTPP is flowed into CCUS, then captured CO2 is provided to Sabatier reactor to react with H$_2$ electrolyzed to produce CH$_4$. The external characteristics are adopted in planning problem instead of the internal chemical reaction process [36], Fig. 2 shows the flow chart retrofitting TTPP to CCPP with CCUS.

$$P_{j,t} = P_{j,t}^{CCPP} + P_{j,t}^{PtG} + P_{j,t}^{CC}, \forall j \in \Psi^{GEN}, \forall t \in [1,T] \quad (21)$$

$$Q_{j,t}^{emi} = emi \cdot P_{j,t}, \forall j \in \Psi^{GEN}, \forall t \in [1,T] \quad (22)$$

$$Q_{j,t}^{CC} = P_{j,t}^{CC}/W^{CC}, \forall j \in \Psi^{GEN}, \forall t \in [1,T] \quad (23)$$

Eq. (21) denotes the TTPP output power supply for the external grid, PtG, and carbon capture device. In (22), considering that the CO$_2$ emissions of power plants changed with loads, the CO$_2$ emission volume accounts for a fixed proportion $emi$ of output power based on statistical data from U.S. DOE [39]. The relationship between the captured CO$_2$ volume by carbon capture device, i. e. $Q^{CC}$, and unit work by carbon capture device operation power $P^{CC}$ is approximated by linear correlation with $W^{CC}$ in (23), $W^{CC}$ (unit: MWh/ton) is the energy consumption of capturing per ton CO$_2$ by carbon capture device adopted from [37].

$$Q_{j,t}^{CC} = Q_{j,t}^{CS} + Q_{j,t}^{CU}, \forall j \in \Psi^{CCPP}, \forall t \in [1,T] \quad (24)$$

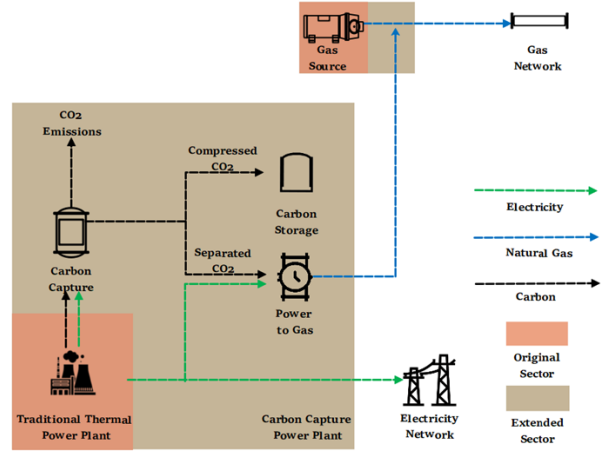

Fig. 2 Framework of CCPP and PtG coupling in EGC-IES

The captured CO$_2$ is either used for PtG to produce methane ($Q^U$) as separated form or stored to sell to carbon market ($Q^{CS}$) in compressed form, which can be modeled by (24).

$$Q_{j,t}^{CU} = \alpha^{CO_2}\eta^{PtG}P_{j,t}^{PtG}, \forall j \in \Psi^{CCPP}, \forall t \in [1,T] \quad (25)$$

$$V_{j,t}^{CH_4} = \eta^{PtG}P_{j,t}^{PtG}/H^{CH_4}, \forall j \in \Psi^{CCPP}, \forall t \in [1,T] \quad (26)$$

$$y_j^{PtG}\underline{P}^{PtG} \leq P_{j,t}^{PtG} \leq y_j^{PtG}\overline{P}^{PtG}, \forall j \in \Psi^{CCPP}, \forall t \in [1,T] \quad (27)$$

The total CO$_2$ volume consumed by PtG during time period $t$ can be calculated as (25), where $\eta^{PtG}$ is the conversion efficiency of PtG from electricity to CH$_4$, $\alpha^{CO_2}$ (unit: ton/MWh) is CO$_2$ consumed volume of PtG to generate unit work CH$_4$ adopted from [38]. Constraints (26) denote CH$_4$ production character of PtG, $H^{CH_4}$ is the calorific value of methane, usually takes 36MJ/m$^3$, $V_{j,t}^{CH_4}$ signifies produced CH$_4$ volume by PtG in CCPP $j$ during hour $t$. Constraints (27) limits the output power of PtG in CCPP $j$ during hour $t$ with boundaries $\overline{P}^{PtG}$ and $\underline{P}^{PtG}$ multiplied by planned module $y_j^{PtG}$.

## E. Constraints on Facility Investment and Siting

$$0 \leq \sum_{m \in GN} s_{m,j} \leq y_j^{PtG}M, \forall j \in CCPP \quad (28)$$

$$0 \leq V_{m,j,t}^{CH_4} \leq s_{m,j}M, \forall m \in \Psi^{GN}, \forall j \in \Psi^{CCPP}, \forall t \in [1,T] \quad (29)$$

$$V_{j,t}^{CH_4} = \sum_m V_{m,j,t}^{CH_4}, \forall m \in \Psi^{GN}, \forall j \in \Psi^{CCPP}, \forall t \in [1,T] \quad (30)$$

Natural gas generated by PtG can be delivered to the gas network to relieve gas supply pressure. $s_{m,j}$ is the binary variable indicating the investment statue of transmission pipeline between gas node $m$ and PtG $j$, which is equal to 1 if they are connected, otherwise it is 0. $y_j^{PtG}$ is integer variable denoting the number of PtG module to be invested, with each module assumed to be 1MW. Constraint (28) means that every pipeline can only be invested after PtG was chosen to be expanded on CCPP due to investment logic, where $M$ is a very large number used in the Big-M method. Similarly in (29), the transmission volume of natural gas between gas node $m$ and PtG $j$ at any time is bound by decision variable $s_{m,j}$. (30) implicates the coupling relationship between natural gas produced by PtG and transmitted in pipeline. Fig. 3 illustrates the sketch on facility sizing and siting in EGC-IES.

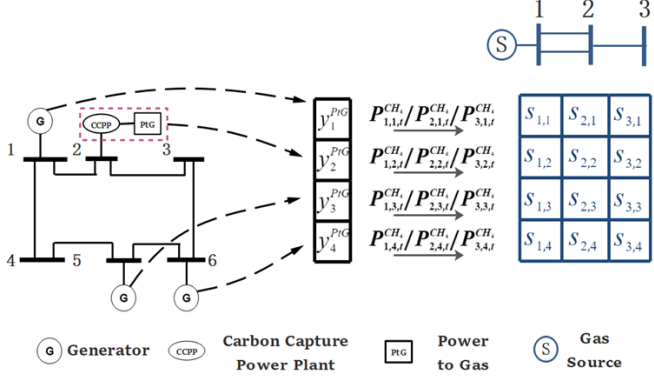

Fig. 3 Sketch on facility investment and siting of EGC-IES

$$\mathbf{A}_{m,i}P_{i,t}^{G} + \mathbf{B}_{m,p}f_{p,t}^{gas} + \sum_j V_{m,j,t}^{CH_4} = L_{m,t} \quad (31)$$

$\forall m \in \Psi^{GN}, \forall i \in \Psi^{GS}, \forall p \in \Psi^{GP}, \forall j \in \Psi^{CCPP}, \forall t \in [1,T]$

$$\mathbf{C}_{n,j}P_{j,t}^{CCPP} + \mathbf{D}_{n,l}f_{l,t}^{ele} = L_{n,t} \quad (32)$$

$\forall n \in \Psi^{EB}, \forall j \in \Psi^{CCPP}, \forall l \in \Psi^{TL}, \forall t \in [1,T]$

After retrofitting TTPP to CCPP with CCUS and PtG, the original power balance functions in (12) and (20) are converted into (31) and (32) by adding an external gas source from PtG and replacing generator as CCPP.

## IV. METHODOLOGY

In Section IV, three models are formulated based on previous constraints and objective function, and the effectiveness of proposed models are verified through case studies in Section V.

### A. Without CCUS planning

Without CCUS planning is the original circumstance, in which it does not exist additional investment in CCUS equipment. The $CO_2$ generated by TTPP is directly emitted into the atmosphere, and the utilities have to pay the carbon penalty. Correspondingly, there is no carbon revenue due to the absence of CCUS.

The objective function in (28) consists of operation costs of gas sources, generators, and carbon penalty due to $CO_2$ emission. It could be put into a compact form:

$$\min\left\{\sum_T \left[C_{ope}^{GS}(P_{i,t}) + C_{ope}^{GEN}(P_{j,t}) + C_{pen}(P_{j,t})\right]\right\},$$

$\forall i \in \Psi^{GS}, \forall j \in \Psi^{GEN}, \forall t \in [1,T]$ (33)

$s.t. (6)-(20),(22)$

### B. Deterministic planning with CCUS

Deterministic planning with CCUS considers investment cost, operation cost, carbon capture/storage cost and carbon penalty/revenue in a specific value of carbon price and carbon tax. Different planning results can be obtained since the carbon price and carbon tax fluctuations along with policy changes. It could be put into a compact form:

$$\min\left\{\begin{array}{l}\kappa^{PtG}c_{inv}^{PtG}(y_j^{PtG}) + \kappa^{siting}c_{inv}^{siting}(s_{m,j}) + \\ \sum_{t=1}^{T}\left[\begin{array}{l}C_{ope}^{GS}(P_{i,t}) + C_{ope}^{GEN}(P_{j,t}) + C_{ope}^{PtG}(P_{q,t}) \\ +C^{CC}(P_{j,t}) + C^{CS}(P_{j,t}) + C_{pen}(P_{j,t}) + C_{tra}(P_{j,t})\end{array}\right]\end{array}\right\} \quad (34)$$

$s.t. (6)-(11b),(13a)-(19),(21)-(32)$

$\forall i \in \Psi^{GS}, \forall j \in \Psi^{CCPP}, \forall q \in \Psi^{PtG}, \forall t \in [1,T]$

### C. Two-stage stochastic planning with CCUS

There are over 61 carbon pricing initiatives in place or scheduled for implementation until 2020 [4], meanwhile, the carbon price level of the implemented carbon pricing mechanism varies greatly around the world. Therefore, it is essential to consider the carbon price and carbon tax uncertainty based on the implemented carbon pricing mechanism globally. In the two-stage stochastic planning model, the first-stage decisions include the PtG capacity and its siting, while the second-stage considers the operation cost, e. g. carbon capture & storage cost, carbon penalty & trading in all scenarios.

$$\min\left\{\begin{array}{l}\kappa^{PtG}c_{inv}^{PtG}(y_j^{PtG}) + \kappa^{siting}c_{inv}^{siting}(s_{m,j}) + \\ \sum_s P_s\left[\begin{array}{l}C_{ope}^{GS}(P_{i,t}(s)) + C_{ope}^{GEN}(P_{j,t}(s)) + C_{ope}^{PtG}(P_{q,t}(s)) \\ +C^{CC}(P_{j,t}(s)) + C^{CS}(P_{j,t}(s)) + C_{pen}(P_{j,t}(s)) + C_{tra}(P_{j,t}(s))\end{array}\right]\end{array}\right\}$$

EXPECTED VALUE

$s.t. (6)-(11b),(13a)-(19),(21)-(32)$

$\forall i \in \Psi^{GS}, \forall j \in \Psi^{CCPP}, \forall q \in \Psi^{PtG}, \forall t \in [1,T]$

(35)

### D. Two-stage robust planning with CCUS

Unlike the method using probabilistic weights of uncertainties in stochastic planning, two-stage robust planning further considers the robustness of uncertainty set in the second stage. Box uncertainty set is adopted to illustrate the uncertainty of policy considering carbon price and carbon tax, the boundaries of uncertainty set are obtained from recent *Prices in implemented carbon pricing initiatives* in [5].

$$\min\left\{\begin{array}{l}\kappa^{PtG}c_{inv}^{PtG}(y_j^{PtG}) + \kappa^{siting}c_{inv}^{siting}(s_{m,j}) + \\ \max_{(r^{tax},r^{pr})\in U}\min\sum_T\left[\begin{array}{l}C_{ope}^{GS}(P_{i,t}) + C_{ope}^{GEN}(P_{j,t}) + C_{ope}^{PtG}(P_{q,t}) \\ +C^{CC}(P_{j,t}) + C^{CS}(P_{j,t}) + C_{pen}(P_{j,t}) + C_{tra}(P_{j,t})\end{array}\right]\end{array}\right\}$$

$U = \left\{r^{tax},r^{pr} \middle| r^{tax} \in \left[\underline{r}^{tax},\overline{r}^{tax}\right], r^{pr} \in \left[\underline{r}^{pr},\overline{r}^{pr}\right]\right\}$

$s.t. (6)-(11b),(13a)-(19),(21)-(32)$

$\forall i \in \Psi^{GS}, \forall j \in \Psi^{CCPP}, \forall q \in \Psi^{PtG}, \forall m \in \Psi^{GN}, \forall n \in \Psi^{EN}, \forall t \in [1,T]$

(36)

## V. CASE STUDIES

An updated IEEE 24-bus electric system with the Belgian 20-node natural gas system are employed to verify the effectiveness of the proposed models. A sketch map indicating sources of key parameters of PtGs and their siting planning is shown in Appendix Table. All the datasets are available in [40]. The investment cost of PtG is set as 3 M$/MW, the siting costs of PtG to gas system are set to different values according to distance. The MILP is modeled by MATLAB 2019b with YALMIP toolbox and solved by Gurobi 9.1 on a laptop with Intel® Core™ i7-6700U 3.40 GHz processor and 8GB RAM.

### A. Analysis of planning results

Four cases are developed to observe corresponding planning and operation results based on the proposed models as follows:

*1) Case 1*

Operation results without CCUS planning ($r^{tax}$ =50 $/ton);

*2) Case 2*

Deterministic planning results with CCUS($r^{tax}$ =50 $/ton, $r^{pr}$ =40 $/ton);

*3) Case 3*

Two-stage stochastic planning results with CCUS in 25 scenarios (the value of $r^{tax}$ and $r^{pr}$ are selected as five-segment points on average within their range to form 5*5 equal-probabilistic scenarios, the range of $r^{tax}$ is $[1,120]$, the range of $r^{pr}$ is $[1,80]$);

*4) Case 4*

Two-stage robust planning results with CCUS($r^{tax}$ and $r^{pr}$ are uncertain variables, $r^{tax} \in [1,120]$, $r^{pr} \in [1,80]$).

Table II Summary of Compared Cases

| Case | With CCUS | Deterministic planning | Stochastic planning | Robust planning |
|---|---|---|---|---|
| Case 1 | × | × | × | × |
| Case 2 | √ | √ | × | × |
| Case 3 | √ | × | √ | × |
| Case 4 | √ | × | × | √ |

Table II summarizes the problem identifications of the proposed four cases. The detailed planning results are shown in Table III.

Overall, the investment cost increased from Case 2 to Case 4. The reasons can be traced back to the second stage, the second stage of Case 2 is calculated in a specific scenario ($r^{tax}$=50 $/ton, $r^{pr}$=40 $/ton); while the second stage of Case 3 is the weighted average based on 25 equal-probabilistic scenarios extracted from the current carbon price and carbon tax range globally, with more scenarios simulated, the planning result is likely to get closer to the average value ($r^{tax}$=60 $/ton, $r^{pr}$=40 $/ton); as for Case 4, the planning result reflects the optimal decisions in the worst scenario of the current global carbon policy ($r^{tax}$=120 $/ton, $r^{pr}$=1 $/ton), where the extreme situation probably won't happen in practice. The siting planning of CCUS obtained the same results, gas node #17 is selected as the connection node of gas network and electricity network due to joint optimization based on siting cost and load conditions of gas network node. More detailed analysis of investment costs can be found in Section V.B.

As for the operation cost, the gradual rise of generators' and PtGs' indicate their outputs increased along with the carbon capture process and PtGs' increasing investment. The operation costs of gas sources decreased slightly while the generators generate more power (174.60 in Case 1 →334.62 in Case 4), revealing that gas sources' supply pressure is eased with the coupling PtGs producing gas from CCUS in EGC-IES.

Carbon-related indices are also our key concerns. Case 1 (without CCUS planning) accounted for maximum carbon emission volume and carbon penalty except for Case 4 (robust planning with the highest carbon tax), the carbon emission volume increases from Case 2 to Case 4 since outputs of generators to supply for CCUS is also increased to avoid more penalty due to higher carbon tax, while the carbon capture and storage volume of Case 2 and Case 3 varied little. In Case 4, with the highest carbon tax and lowest carbon price, we obtained the worst circumstance with maximum carbon penalty (1588.40M$) and minimum carbon revenue (15.04M$) of the EGC-IES.

In summary, for the global average carbon price and carbon tax level, the business model of installing CC-CS-PtG CCUS for TTPP and transmitting produced gas to the gas network is economically feasible. However, for those countries or regions with high carbon taxes and low carbon prices, the business model of retrofitting such kind of CCUS for policy arbitrage is still not applicable. The next subsection analyzed the approximate range of carbon prices and carbon taxes that have economic advantages in terms of the total cost.

Table III. PLANNING RESULTS

| Category | Indices | Case 1 | Case 2 | Case 3 | Case 4 |
|---|---|---|---|---|---|
| Annualized Investment Cost (M$) | CCUS * | - | 36.61 | 72.58 | 428.69 |
| | CCUS sittings | - | 50.74 | 50.74 | 50.74 |
| | Total | - | 87.35 | 123.32 | 479.43 |
| Operational Cost (M$) | Gas Sources | 4809.61 | 4761.19 | 4760.28 | 4751.27 |
| | Generators | 174.60 | 288.44 | 292.56 | 334.62 |
| | PtGs | - | 0.90 | 1.79 | 10.58 |
| | Capture | - | 723.68 | 705.26 | 470.37 |
| | Storage | - | 120.34 | 117.00 | 75.22 |
| | Penalty | 967.89 | 110.55 | 186.31 | 1588.40 |
| | Revenue | - | 962.74 | 936.05 | 15.04 |
| | Total | 5952.10 | 5042.37 | 5127.17 | 7215.44 |
| Total Cost (M$) | | 5952.10 | 5129.72 | 5250.49 | 7694.87 |
| Carbon-related Volume (Mt) | Emission | 19.36 | 2.21 | 3.10 | 13.23 |
| | Capture | - | 24.12 | 23.51 | 15.68 |
| | Storage | - | 24.06 | 23.40 | 15.04 |
| | Utilization | - | 0.06 | 0.11 | 0.64 |

* Contains CC, CS and PtG device

### B. Sensitivity analysis of carbon price and carbon tax

The investment decisions in different carbon prices and carbon taxes are simulated in Fig. 4 changed in steps of $10. It can be concluded from the figure that the investment tends to rise as carbon price decreases and carbon tax increases. It is reasonable under the current business model: the optimal choice is to sell the captured $CO_2$ for income when the carbon price is rather high; with relatively low carbon price, it is not cost-effective to sell the captured $CO_2$ and it would be better to produce methane to reduce gas supply pressure. Furthermore, it can be observed from Fig. 4 that when carbon price is higher

than carbon tax (lower right corner of Fig. 4), no more CCUS investment is preferred; on the contrary, when carbon price is lower than carbon tax, CCUS investment is adopted, and the investment increases with difference increases. The results depict that CCUS retrofitting is an investment strategy sensitive to the carbon emission policy, i.e., carbon tax and carbon price.

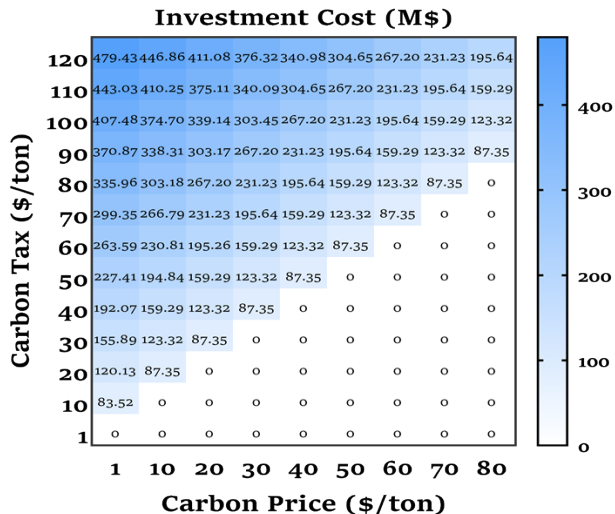

Fig. 4 Investment cost in different carbon tax and carbon price

Fig. 5 is to further analyze the trend of total cost along with carbon price and carbon tax compared to circumstance without CCUS planning, in which surface in pink color represents the total cost without CCUS. Generally speaking, the total cost presents a downward tendency along with the carbon price increases and carbon tax decreases. It can be seen from the figure that while the value of carbon price is larger than 40 $/ton, the business model of CCUS retrofitting with CC, CS and PtG for TTPP is approachable globally. This retrofit planning has achieved economically better for some high-carbon tax areas while the carbon price equals 30 $/ton.

It is worth mentioning that, despite carbon prices increasing in many jurisdictions all over the world, the carbon prices in most parts of the world are still below 40$/ton at present. Report [5] has pointed out that the carbon price level of 50$/ton-100$/ton by 2030 is required to cost-effectively reduce emissions in line with the temperature goals in the *Paris Agreement*, this is also consistent with the intersecting line shown in Fig. 5. Facing the policy foundation of a gradual increase in carbon prices, the retrofit planning could have great potential in the future.

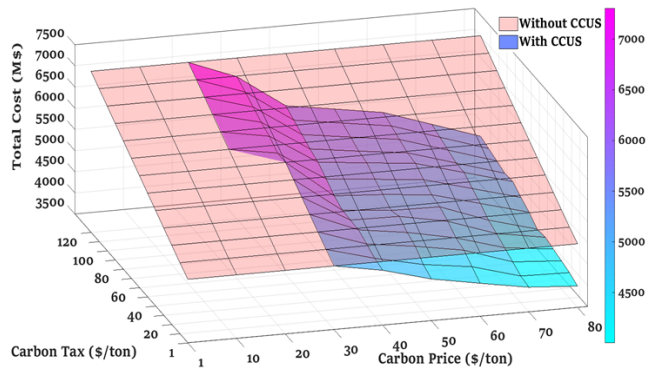

Fig. 5 Total cost in different carbon tax and carbon price

### C. Analysis of daily carbon indices

The effectiveness of the proposed models can be verified further by comparing daily carbon emission, capture, storage, and utilization volume in Appendix Fig. Case 1 in blue color can be regarded as a benchmark without CCUS planning, it results in vast amounts of carbon emissions and simultaneously wastes policy dividends. Case 2 and Case 3 can be compared and analyzed together, in which more sampled scenarios can lead closer to the global average level in Case 3. Case 4 in purple line reflects the "worst case" in robust optimization: compared with Case 2 and Case 3, the robust planning result is more inclined to carbon emission and averse to carbon capture/storage due to the extremely low carbon price, even when carbon tax is also high; at the same time, the planning result adopted maximum carbon emissions volume and minimum carbon capture/storage volume in the worst circumstance. Moreover, captured carbon is used for storage or utilization, they made a tradeoff with each other. Because of the minimum operating power constraint with the planned capacity of PtG (Constraints (27)), the carbon utilization volume in Case 4 still achieved the highest among all cases.

Carbon-related volume is also consistent with the load conditions. During the morning and evening load valley moment (1:00-6:00, 22:00-24:00), CCUS is sufficient to achieve zero carbon emissions; the changing trend of carbon capture and carbon storage curves is basically in accordance with load conditions for Case 2 and Case 3. During non-valley periods (7:00-22:00), abnormal changes of carbon capture and storage happened due to the worst carbon price in Case 4.

On the whole, it can be observed that the effectiveness of the proposed models (deterministic planning, two-stage stochastic planning & two-stage robust planning) in reducing carbon emissions, increasing carbon capture and utilization by making use of carbon policy dependent on carbon tax and prices.

### VI. CONCLUSIONS

This paper presents a gas-electricity coupled integrated energy system planning model with CCUS, in which carbon price and tax uncertainties are included. With proper carbon tax and price range, the proposed model effectively reduces carbon emissions, increases carbon capture and usage, and makes use of policy to profit. In the proposed model, the capacity and siting of PtG are included in the objective function to optimize the first-stage investment cost, and the economic indices on operation cost of generator, gas source and PtG, carbon capture and storage cost, as well as carbon penalty and revenue are optimized by IES operation strategy in the second stage.

Moreover, carbon price and carbon tax were considered in the model as the key sources of the uncertainties under the changing carbon policy. Stochastic and robust planning methods are introduced to verify mutually through economic indices and carbon indices. The case studies demonstrate the effectiveness of the proposed gas-electricity coupled models by PtG and illustrated the benefits of CUSS installation. Future work could further consider gas turbine and power to gas to form bi-direction energy flows and renewable energy and load uncertainties in EGC-IES.

APPENDIX

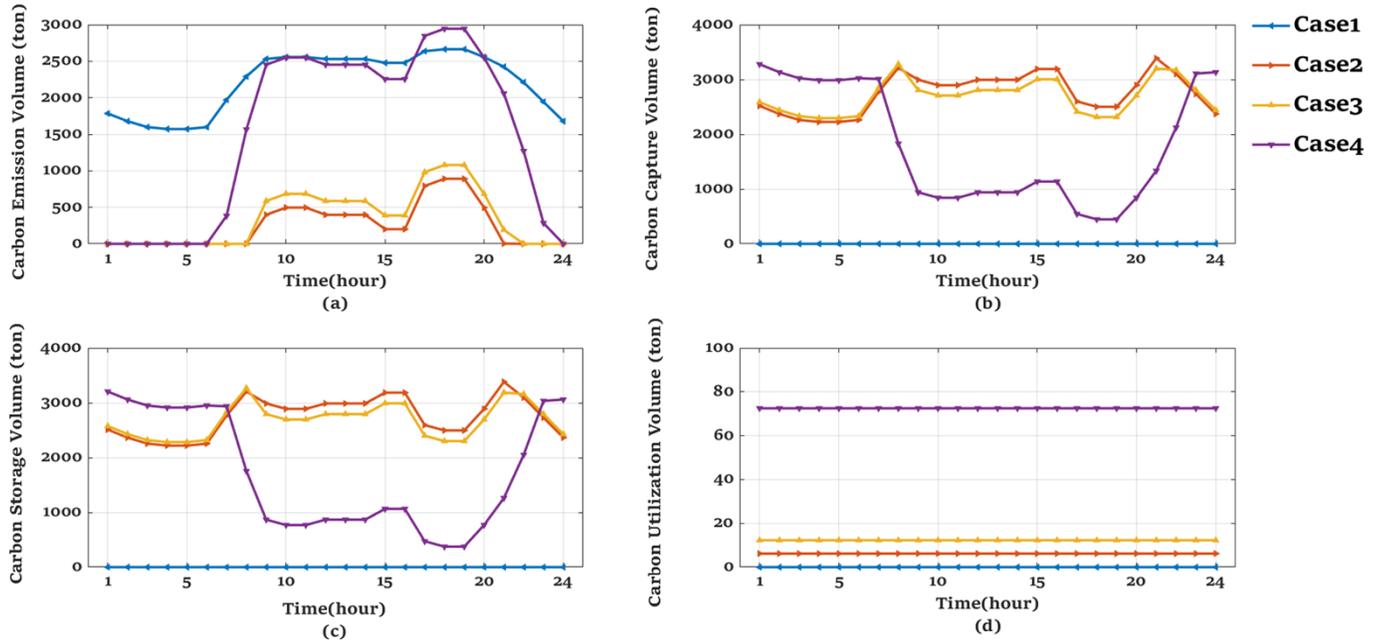

Appendix Fig. Daily carbon emission, capture, storage, and utilization volume
(a). carbon emission volume. (b). carbon capture volume.
(c). carbon storage volume. (d). carbon utilization volume.

Appendix Table. Source of Key Parameters

| Parameter | Value | Source |
|---|---|---|
| $CO_2$ emission factor of TTPP $emi$ | 1005 g/KWh | Website [39] |
| Carbon capture cost | ≤30 USD/ton | Website [4] |
| Carbon storage cost | ≤10 USD/ton | |
| Carbon tax | 1~120 USD/ton | Report [5] |
| Carbon prize | 1~80 USD/ton | |
| CCPP energy consumption of capturing per ton $CO_2$ $W^C$ | 0.269 MWh/ton | Literature [37]-[38] |
| $CO_2$ consumed volumn of PtG per unit output $\alpha^{CO_2}$ | 0.2 ton/MWh | |
| Conversion efficiency of PtG $\eta^{PtG}$ | 0.6 | |